# Regions, Innovation Systems, and the North-South Divide in Italy

Loet Leydesdorff,[a] * & Ivan Cucco[b]

**Abstract**

Using firm-level data collected by Statistics Italy for 2008, 2011, and 2015, we examine the Triple-Helix synergy among geographical and size distributions of firms, and the NACE codes attributed to these firms, at the different levels of regional and national government. At which levels is innovation-systemness indicated? The contributions of regions to the Italian innovation system have increased, but synergy generation between regions and supra-regionally has remained at almost 45%. As against the statistical classification of Italy into twenty regions or into Northern, Central, and Southern Italy, the greatest synergy is retrieved by considering the country in terms of Northern and Southern Italy as two sub-systems, with Tuscany included as part of Northern Italy. We suggest that separate innovation strategies should be developed for these two parts of the country. The current focus on regions for innovation policies may to some extent be an artifact of the statistics and EU policies. In terms of sectors, both medium- and high-tech manufacturing (MHTM) and knowledge-intensive services (KIS) are proportionally integrated in the various regions.

[a] *corresponding author; Amsterdam School of Communication Research (ASCoR), University of Amsterdam, PO Box 15793, 1001 NG Amsterdam, The Netherlands; loet@leydesdorff.net
[b] Graduate School, The American University of Rome, Via Pietro Roselli 4, 00153 Roma RM, Italy; i.cucco@aur.edu

## 1. Introduction

Italy was shaped as a modern nation state in the period 1860-1870. During the war of 1860-1861, the northern part was unified under the leadership of the Kingdom of Piemonte (Turin), and the southern part—the Kingdom of the Two Sicilies (with Naples as capital)—was conquered by Garibaldi in that same year. Central Italy, which until then had been the Papal State, was invaded by Italy in 1870 and thereafter Rome became the capital of the nation. The division into three parts—Northern, Central, and Southern Italy—has, however, remained important; it is commonly used for analytical and policy purposes. However, the North/South divide is also a common terminology in political discourse: the "questione meridionale" or the Southern Question. In short, the North and the South have different cultural traditions and marked differences in GDP per capita, composition of economic activities, and employment indicators.

At a lower level of aggregation, the country is administered in terms of twenty regions of which three are semi-autonomous: Sicilia, Sardegna, and Friuli Venezia Giulia. Valle d'Aosta is an autonomous region, in which French functions as a second language, while Alto Adige (also known as Süd-Tirol) is an autonomous province of the region Trentino-Alto Adige bordering on Austria, with German as a second language. Below the level of regions, 107 provinces are defined. Furthermore, Italy is well known for its "industrial districts" which often cover a small territory within one or more provinces, with specialized manufacturing or services (Becattini *et al.*, 2003; Bertamino *et al.*, 2017). These districts are highly innovative and mainly located in the northern part of the country (Biggiero, 1998). However, they are not a separate level of administration and hence not included in the national statistics.[1] The latter are aligned with the hierarchical classification of the European Union in the "Nomenclature des Unités Territoriales Statistiques" (Nomenclature of Territorial Units for Statistics, or NUTS). In the NUTS classification, NUTS1 is defined as lands (e.g., the German *Länder*), NUTS2 as regions (e.g., Lombardia), and NUTS3 as provinces or metropolitan cities (e.g., the metropolitan region of Milano or the province of Lecce).

Grilliches (1994) noted that the use of administrative units in statistics can be a data constraint for innovation studies and also for innovation policies. For example, innovation is not geographically constrained (Carlsson & Stankiewicz, 1991). Innovation systems may depend on interactions and infrastructures that do not match regional and national boundaries. Sectorial innovation systems (e.g., oil refinery; biotechnology) are in important respects organized internationally (Carlsson, 2006). Furthermore, it has been shown that firms interact with non-regional universities if the knowledge and skills required are not available within the region (Asheim & Coenen, 2006; Fritsch & Schwirten, 1999) or when they are seeking higher quality collaboration partners at the international level (d'Este & Iammarino, 2010; Laursen, Reichstein, & Salter, 2011).

In a recent study of the U.S. innovation system, Leydesdorff *et al*. (2018, in preparation) found that the regions which are measured in the U.S.A. as Core-Based Statistical Areas (CBSA) are often too small to comprise innovation systems; the innovation systems spill over the boundaries of these units of analysis. An alternative would be to focus on groups of contiguous CBSAs, but the analysis is then no longer supported by the national and regional organization of statistics.

---

[1] Using 2011 census data, Statistics Italy identified 611 local labour systems ("sistemi locali del lavoro", SLL) based on commuting patterns. Many of these areas overlap with industrial districts, and therefore allow for economic analyses at the district level (e.g., Paci & Usai, 1999; Mameli, Faggian, & McCann, 2008). Comparable SLL-level data on employment by sector are available for the years 2001 and 2011. However, this data is not available at the micro-level of individual firms.

For the purpose of implementing innovation policies at the appropriate level, it is nevertheless important to understand the boundaries of innovation systems. This is a complex undertaking which could be addressed at different levels (e.g. municipal, provincial, regional, national, supra-national; by sector or comprehensively) and using different instruments, such as various combinations of qualitative analyses and batteries of quantitative indicators. In this study, we focus on Italy as a challenging and exemplary case: to what extent and at which level is innovation-systemness indicated? Can the regions carry the function of regional innovation organizers (Etzkowitz & Klofsten, 2005)? We shall argue that the current understanding of Italy in terms of regional and supra-regional innovation systems is not optimal in terms of the possible synergies at regional and national levels among (*i*) the geographical distributions of firms, (*ii*) the economic structure in terms of firm sizes, and (*iii*) the technological knowledge bases of these firms as indicated by the NACE-codes. (NACE is the abbreviation for the "Nomenclature générale des Activités économiques dans les Communautés Européennes" used by the OECD and EuroStat.)

Storper (1997) has called the quality of the relations among these three dimensions—geography, technology, and organization—"a holy trinity." This accords with the perspective of a Triple Helix of university-industry-government relations in which the dynamics of knowledge, economics, and control are combined. Synergy in these relations does two things: it reduces uncertainty and generates new options. Reduction of uncertainty can be expected to improve the climate for investments (Freeman & Soete, 1997, pp. 242 ff.); new options provide opportunities for the survival of new activities in the highly competitive markets of emerging technologies (Bruckner, Ebeling, Montaño, & Scharnhorst, 1996).

## 2. Innovation systems and innovation policies

The concept of *national* innovation systems was first proposed by Freeman (1987) as a possible "lesson from Japan." In the years thereafter, Lundvall (1993) and Nelson (1993) provided two collections of comparative studies among nations. However, the emphasis on "national" more or less provoked the question of whether innovation systems might also be regional. On the one side, regions such as Catalonia, Flanders, and Wales have autonomous aspirations. At the level of the European Union, on the other side, the metaphor of an emerging "knowledge-based economy" rapidly became more popular than a focus on individual nations (Foray & Lundvall, 1996; Commission of the European Community, 2000).

Both the OECD and the EU provide strong incentives for organizing regional innovation agencies and programs. Among other things, the OECD reviews regional innovation policies with the objective of providing policy recommendations (e.g., OECD, 2009). In innovation studies (economic geography and evolutionary economics), it is increasingly assumed that regions (including metropolitan regions) are the appropriate units of analysis for studying the transition to a knowledge-based economy (e.g., Braczyk, Cooke, & Heidenreich, 1998; Cooke,

2002; Feldman & Storper, 2016; ; Florida, 2002; Storper, Kemeny, Makarem, Makarem, & Osman, 2015).)

In Italy, regions have gained importance as innovation-policy units since 2001, when a range of devolution measures gave regional governments greater control over policy areas such as health, education, and economic and industrial development, including innovation policy (Rolfo & Calabrese, 2006). This devolution led to a sharp reduction of the national budget for the support of industrial and R&D activities, particularly in the South. Brancati (2015) estimates that between 2002 and 2013, state aid decreased by 72%; the remaining state interventions privileged Central and Northern Italy, while industrial policies in favor of the Southern regions were virtually abandoned after 2000 (Prota & Viesti, 2013).

Against this backdrop, the 2007-2009 economic and financial crisis has severely impacted the Italian industrial system. Compared with the trends calculated for the 1992-2008 period, about 300 bn Euro of gross investment were lost in Italy between 2008 and 2013 (Cappellin *et al.*, 2014). Southern regions were disproportionally affected: between 2007 and 2012, industrial investment in the South decreased by 47% (Prota & Viesti, 2013). This retreat of national policy has only partly been compensated by regional policies, supported to varying degrees by EU Cohesion and Structural funds. In the EU programs during the period 2007-2013, about 21.6 bn Euro of EU funds (FESR/ERDF and FSE/EFS) were allocated to regions in Southern Italy for Convergence objectives (Calabria, Campania, Puglia, and Sicilia) and 6.3 bn to regions in Central and Northern Italy for Competitiveness objectives.

Despite the increasing role played by regional governments in innovation policy, it has remained a subject of debate whether the regional level is most appropriate for the design and implementation of such policies. On the basis of an analysis of the performance of the Italian national innovation system during the 1980s and 1990s, Malerba (1993, at p. 230), for example, argued that "not one, but two innovation systems are present in Italy." The first one is a "core R&D system" that operates at the national level through systematic cooperation between large firms with industrial laboratories, small high-tech firms, universities, public research institutes, and the national government. The second innovation system would be a "small-firms network" composed of a plurality of small- and medium-sized firms that cooperate intensively at the local level, often within industrial districts, and generate incremental innovation through learning-by-doing.

Malerba mentions the lack of overall coordination in public policy and R&D support services and a weak tradition of successful university-industry cooperation in research as major problems in the Italian innovation system. Nuvolari & Vasta (2015) added that Italy can be characterized as a structurally weak national innovation system in comparison to its main competitors. The diverging performance between scientific and technological activities can lead to major difficulties in the technology transfer of scientific results from universities to firms due to a lack of bridging institutions (e.g., Balconi *et al*., 2004).

A number of studies in various sectors of the economy (e.g., Antonioli *et al*., 2014; Belusssi *et al.,* 2010; De Marchi & Grandinetti, 2017; Lew *et al*., 2018) have argued that the international orientation of research collaborations means that Italian regions cannot be considered as innovation system. These innovative regions are better characterized as "glocal" systems. They pair a relatively low connectedness at the local level with strong knowledge-intensive relationships at the international level. On the industrial side, this international orientation carries a threat of de-industrialization of innovative districts and regions because new options can easily be bought and relocated elsewhere by multinational corporations (Cooke & Leydesdorff, 2006; Dei Ottati, 2003).

In sum, the gradual emergence of knowledge production as an additional coordination mechanism in an industrial system that is otherwise coordinated in terms of institutions and markets introduces the risk of "footloose-ness" (Vernon, 1979). Knowledge-intensive services and high-tech manufacturing uncouple an innovation system from a geographical address and can thus be counter-productive from the perspective of regional innovation policies. Footloose-ness is negative to the local synergy (Leydesdorff & Fritsch, 2006).

## 3. Methods

We operationalize synergy as a reduction of uncertainty in terms of Shannon's (1948) information theory. Using this theory, uncertainty in the distribution of a random variable *x* can be defined as $H_x = -\sum_x p_x \log_2 p_x$. The values of $p_x$ are the relative frequencies of *x*: $p_x = f_x / \sum_x f_x$. When base two is used for the logarithm, uncertainty is expressed in bits of information.

The uncertainty in the case of a system with two variables can be formulated analogously as

$$H_{xy} = -\sum_x \sum_y p_{xy} \log_2 p_{xy} \tag{1}$$

In the case of interaction between the two variables, the uncertainty in the system is reduced by mutual information $T_{xy}$ as follows:

$$T_{xy} = (H_x + H_y) - H_{xy} \tag{2}$$

One can derive (e.g., McGill, 1954, pp. 99 ff.; Yeung, 2008, pp. 59f.) that in the case of three dimensions, mutual information corresponds to:

$$T_{xyz} = H_x + H_y + H_z - H_{xy} - H_{xz} - H_{yz} + H_{xyz} \tag{3}$$

Eq. 3 can yield negative values and is therefore not a Shannon-type information (Krippendorff, 2009). Shannon-type information measures variation, but this negative entropy is generated by next-order loops in the communication, for example, when different codes interact as selection environments.

In other words, when three dimensions operate, uncertainty can be added or reduced by generating mutual information or redundancy, respectively. Additional redundancy reduces relative uncertainty by adding options to the system that were hitherto not realized. Increasing the number of options for further development may be more important for the viability of an innovation system than the options realized hitherto (Fritsch, 2004; Petersen, Rotolo, & Leydesdorff, 2016).

Note that uncertainty is implicated by the variation in *relations*. From an evolutionary perspective, the historical networks of relations function as retention mechanisms. Our measure, in other words, does not measure action (e.g., academic entrepreneurship) as input or output, but the investment climate as a structural consequence of *correlations* among distributions of relations. However, the distinction between these structural dynamics in terms of changing selection environments and the historical dynamics of relations is analytical. The two layers reflect each other in the events. Eq. 3 models this trade-off between variation and selection as positive and negative contributions to the prevailing uncertainty. The question of systemness can thus be made empirical and amenable to measurement: when the generation of redundancy prevails over the generation of uncertainty, systemness is indicated.

In the case of groups (subsamples), furthermore, one can decompose the information as follows: $H = H_0 + \sum_G \frac{n_G}{N} H_G$ (Theil (1972, pp. 20f.). The right-hand term ($\sum_G \frac{n_G}{N} H_G$) provides the average uncertainty in the groups and $H_0$ the additional uncertainty in-between groups. Since *T* values are decomposable in terms of *H* values (Eq. 3), one can analogously derive (Leydesdorff & Strand, 2013, at p. 1895):

$$T = T_0 + \sum_G \frac{n_G}{N} T_G \qquad (4)$$

In this formula, $T_G$ provides a measure of synergy at the geographical scale $G$; $n_G$ is the number of firms at this scale, and $N$ is the total number of firms under study. One can also decompose across regions, in terms of firm sizes, or in terms of combinations of these dimensions.

The three dimensions are the (g)eographical, (t)echnological, and (o)rganizational; synergy will be denoted as $T_{GTO}$ and measured in millibits with a minus sign. Because the scales are sample-dependent, we normalize for comparisons across samples as percentages. After normalization, the contributions of regions or groups of regions can be compared. In this design, the between-group term $T_0$ provides us with a measure of what the next-order system (e.g., the nation) adds in terms of synergy to the sum of the regional systems.

A routine with further instructions is available at http://www.leydesdorff.net/software/th4 which generates the synergy values from data which have for this purpose to be organized as comma-separated variables with for each case (that is, firm) a unique identifier, a postal code, a size class, and a NACE code. The results are organized into a file which can be read into programs like SPSS or Excel for further processing. We use SPSS v.22 to generate the maps of regions in Italy on the basis of the synergy values expressed as percentages of contributions to the overall synergy of the Italian system.

## 4. Data and descriptive statistics

Statistics Italy (IStat) collects firm census data every ten years. In a methodologically oriented study, Cucco & Leydesdorff (2014) used the census data from 2000 for a comparison with data in the ORBIS/Amadeus database of Bureau van Dijk. Although the latter sample covered only 402,316 firms as against 4,247,169 firms in the data of Statistics Italy, the results at the regional level were virtually similar (Spearman's $\rho$ >.99; $p$<.001).

In the meantime, complete data for the years 2008, 2011, and 2015 have become available online from the so-called ASIA ("Archivo Statistico delle Imprese Attive") database of Statistics Italy. This database includes all enterprises that performed productive activities for at least six months during the reference year. It does not cover the sectors agriculture, fisheries, and forestry; public administration and non-profit private organizations are also excluded. The data contain 4,514,022 firms in 2008, 4,450,937 firms in 2011, and 4,338,085 in 2015. (The 2000 industrial census data (4,247,169 firms) was organized a bit differently and therefore we use this latter data only qualitatively for the comparison with our quantitative results.)

For a Triple-Helix analysis of synergy, we need three key variables: (1) the administrative location of the firm in the form of its postal address; (2) the NACE code indicating the main technology in the knowledge base of the firm, and (3) the character of the firm in terms of its size indicated as the numbers of employees. These three dimensions have been used in a number of previous studies about the TH in various nations (see Leydesdorff, Ivanova, & Meyer [2018] for a summary).

*4.1. The geographical distribution of firms in Italy*

The administrative division of Italy into Northern, Central, and Southern Italy and twenty regions is visualized in Figure 1 and further specified in Table 1. Among other things, we will test the three conventional partitions of Italy in columns *b*, *d* and *e* of Table 1.

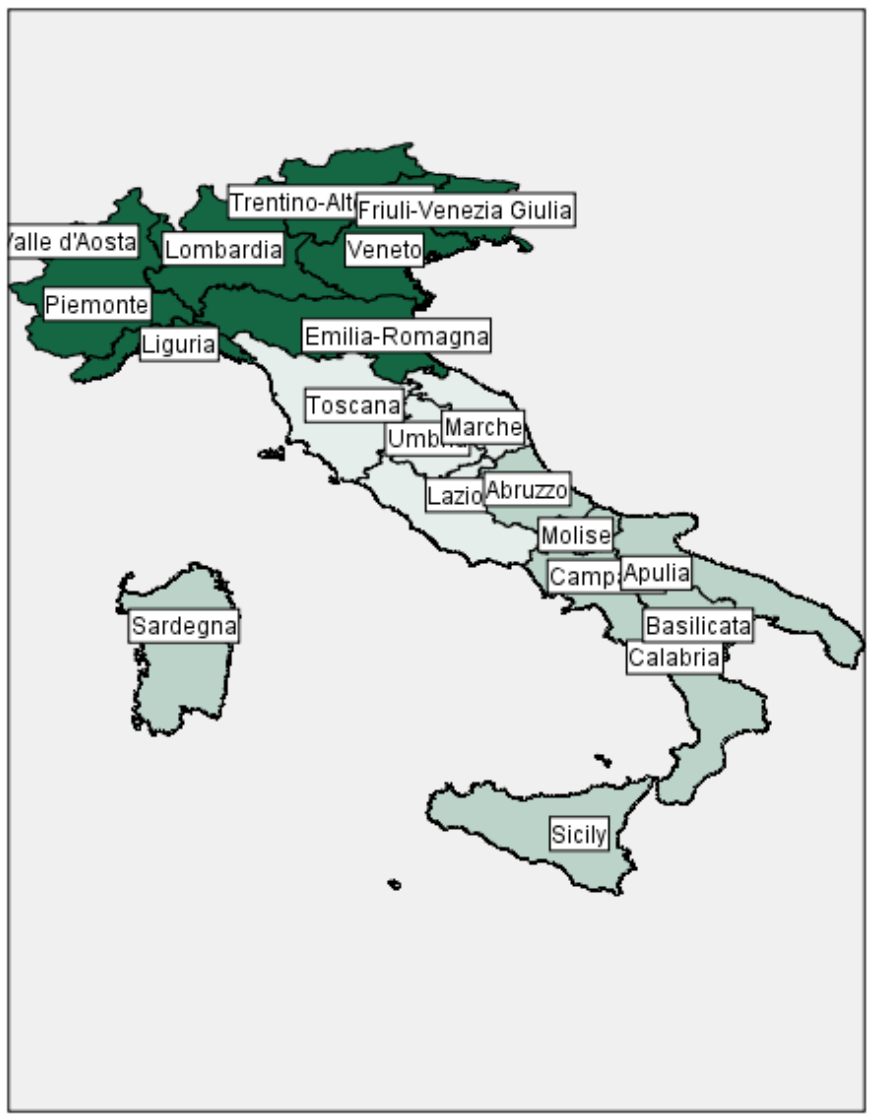

**Figure 1**: Organization of Italy into Northern, Southern, and Central Italy, and regions; Northern Italy is indicated in dark green, Central Italy is in very light green, and Southern Italy is in light green. (Source: figure produced by the authors using SPSS v.22.)

**Table 1**: Regional Division of Italy at the NUTS 1 and NUTS 2 levels.

| Codes of ISTAT | NUTS1 (a) | NUTS2 (b) | Name of the region (c) | Macro-regions (d) | North- South (e) |
|---|---|---|---|---|---|
| 1 | North-west Italy (ITC) | ITC1 | Piemonte | Northern Italy | Northern Italy |
| 2 | | ITC2 | Valle d'Aosta | | |
| 7 | | ITC3 | Liguria | | |
| 3 | | ITC4 | Lombardia | | |
| 4 | North-east Italy (ITH) | ITH1 / ITH2 | Trentino-Alto Adige | | |
| 5 | | ITH3 | Veneto | | |
| 6 | | ITH4 | Friuli Venezia Giulia | | |
| 8 | | ITH5 | Emilia Romagna | | |
| 9 | Central Italy (ITI) | ITI1 | Toscana | Central Italy | |
| 10 | | ITI2 | Umbria | | |
| 11 | | ITI3 | Marche | | Southern Italy |
| 12 | | ITI4 | Lazio | | |
| 13 | Southern Italy (ITF) | ITF1 | Abruzzo | Southern Italy (Mezzogiorno) | |
| 14 | | ITF2 | Molise | | |
| 15 | | ITF3 | Campania | | |
| 16 | | ITF4 | Puglia | | |
| 17 | | ITF5 | Basilicata | | |
| 18 | | ITF6 | Calabria | | |
| 19 | Insular Italy (ITG) | ITG1 | Sicilia | | |
| 20 | | ITG2 | Sardegna | | |

Table 2 provides the numbers of firms in the years under study. In the right-most column we added the 2000 data used by Cucco & Leydesdorff (2014), but since this data was in some respects different we use the previous study only as a point of reference. (For example, Valle d'Aosta was not counted separately in 2000.) The three data points (2008, 2011, and 2015) are sufficient to distinguish trends in the data (Figure 2).

**Table 2**: *N* of firms in 20 Italian regions.*

| Region | 2008 | 2011 | 2015 | (2000)* |
|---:|---:|---:|---:|---:|
| *Piemonte* | 344,334 | 339,261 | 323,184 | 335,749 |
| *Valle d'Aosta* | 11,959 | 11,933 | 11,257 | |
| *Lombardia* | 822,579 | 818,998 | 805,755 | 818,948 |
| *Trentino-Alto Adige* | 83,121 | 83,656 | 84,398 | 82,843 |
| *Veneto* | 406,800 | 402,976 | 391,474 | 405,952 |
| *Friuli-Venezia Giulia* | 88,683 | 86,797 | 82,720 | 89,303 |
| *Liguria* | 132,288 | 129,708 | 122,874 | 132,408 |
| *Emilia-Romagna* | 389,123 | 370,778 | 366,475 | 387,434 |
| *Toscana* | 338,943 | 332,563 | 320,167 | 337,573 |
| *Umbria* | 70,892 | 69,411 | 66,455 | 70,324 |
| *Marche* | 133,261 | 131,567 | 126,213 | 133,942 |
| *Lazio* | 423,059 | 428,715 | 426,322 | 416,460 |
| *Abruzzo* | 100,120 | 101,115 | 97,184 | 100,822 |
| *Molise* | 21,705 | 21,445 | 20,631 | 21,262 |
| *Campania* | 351,688 | 340,601 | 336,819 | 346,337 |
| *Apulia* | 254,431 | 254,277 | 249,196 | 250,264 |
| *Basilicata* | 36,169 | 35,234 | 34,586 | 35,760 |
| *Calabria* | 114,858 | 110,391 | 105,878 | 112,205 |
| *Sicily* | 278,451 | 273,155 | 264,480 | 273,903 |
| *Sardegna* | 111,558 | 108,356 | 102,017 | 108,984 |
| | 4,514,022 | 4,450,937 | 4,338,085 | 4,480,473 |

* The numbers used in the previous study are provided in the right-most column.

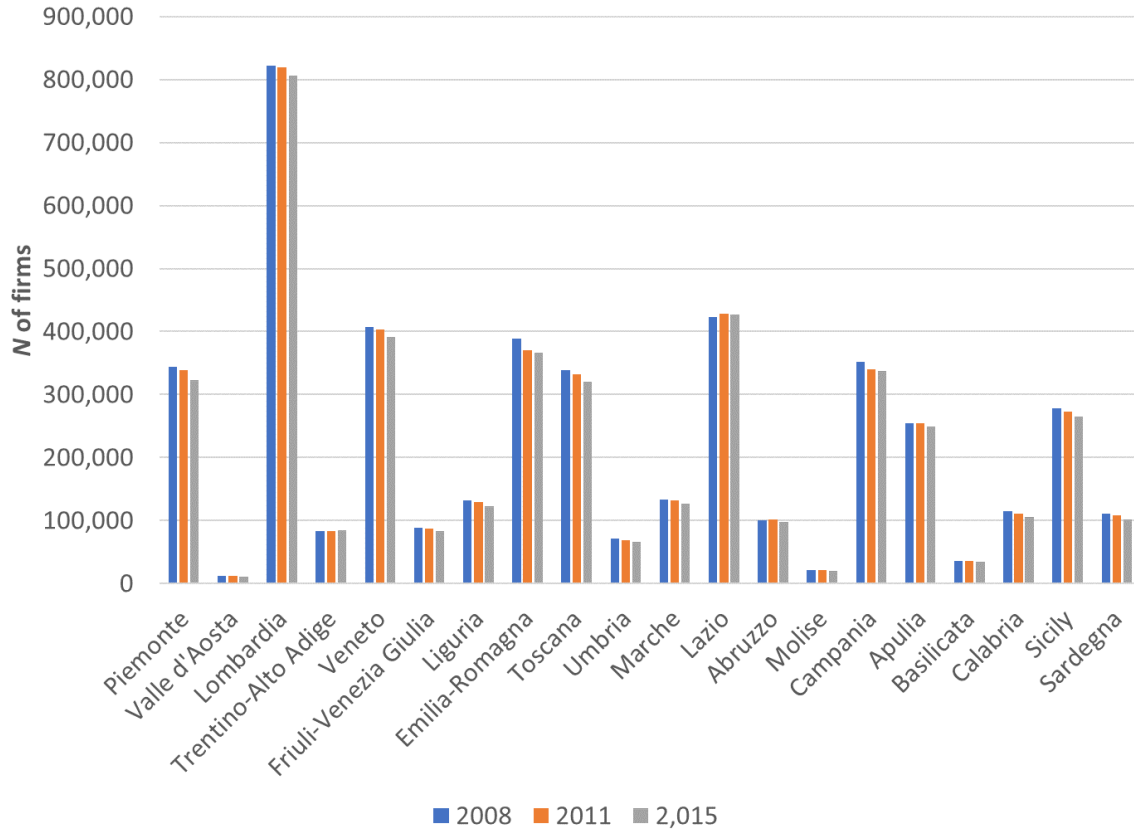

**Figure 2**: *N* of firms in Italian regions in 2008, 2011, and 2015. Source: Statistics Italy.

With the exceptions of Trentino-Alto Adige and Lazio, the numbers of firms have been declining during this past decade. This confirms the impression of stagnation since the crisis of 2008-2009. Italy has only partly recovered from this crisis.

*4.2. Small, medium-sized, and large enterprises*

In addition to the assignment of NACE and postal codes, firms are scaled in terms of the number of their employees. SMEs are commonly defined in terms of this proxy. Financial turn-over is also available in the data as an alternative indicator of economic structure. However, we chose to use the number of employees as one can expect this number to exhibit less volatility than turn-over, which may vary with stock value and economic conjecture more readily than numbers of employees. However, the numbers of employees are sensitive to other activities, such as outsourcing.

The definitions of small and medium-sized businesses, large enterprises, etc., vary among world regions. Most classifications use six or so categories for summary statistics. We use the nine classes provided in Table 3 because this finer-grained scheme produces richer results (Blau & Schoenherr, 1971; Pugh, Hickson, & Hinings, 1969a and b; Rocha, 1999).

**Table 3**: Classification of firms (2015) in terms of the number of employees.
Source: Statistics Italy.

| CLASS | Number of employees | Frequency | Percent | Valid Percent | Cumulative Percent |
|---|---|---|---|---|---|
| 1 | 0 -- 1 | 3,473,928 | 80.1 | 80.1 | 80.1 |
| 2 | 2 -- 4 | 493,365 | 11.4 | 11.4 | 91.5 |
| 3 | 5 -- 9 | 201,497 | 4.6 | 4.6 | 96.1 |
| 4 | 10 -- 19 | 99,554 | 2.3 | 2.3 | 98.4 |
| 5 | 20 -- 49 | 45,476 | 1.0 | 1.0 | 99.4 |
| 6 | 50 -- 99 | 13,275 | .3 | .3 | 99.7 |
| 7 | 100 -- 199 | 6,223 | .1 | .1 | 99.9 |
| 8 | 200 -- 499 | 3,225 | .1 | .1 | 100.0 |
| 9 | 500 or more | 1,542 | .0 | .0 | 100.0 |
|   |   | 4,338,085 | 100.0 | 100.0 |   |

Note that micro-enterprises (with fewer than five employees) constitute 91.5% of the firms under study.

*4.3. NACE codes*

The third dimension of the data to be used is the attribution of NACE codes. The classification of firms in terms of the "Nomenclature générale des Activités économiques dans les Communautés Européennes" (NACE, Rev. 2) is used for indicating the technological dimension.[2] The NACE code can be translated into the International Standard Industrial Classification (ISIC) that is used in the USA (e.g., Leydesdorff, Wagner, Porto-Gomez, Comins, & Phillips, 2018, under submission). The disaggregation in terms of medium- and high-tech manufacturing, and knowledge-intensive services, is provided in Table 4.[3]

**Table 4**: NACE classifications (Rev. 2) of high- and medium-tech manufacturing, and knowledge-intensive services.

| *High-tech Manufacturing* | *Knowledge-intensive Sectors (KIS)* |
|---|---|
| 21 Manufacture of basic pharmaceutical products and pharmaceutical preparations<br>26 Manufacture of computer, electronic and optical products<br>30.3 Manufacture of air and spacecraft and related machinery<br><br>*Medium-high-tech Manufacturing*<br><br>20 Manufacture of chemicals and chemical products<br>25.4 Manufacture of weapons and ammunition<br>27 Manufacture of electrical equipment,<br>28 Manufacture of machinery and equipment n.e.c.,<br>29 Manufacture of motor vehicles, trailers and semi-trailers,<br>30 Manufacture of other transport equipment<br>• excluding 30.1 Building of ships and boats, and<br>• excluding 30.3 Manufacture of air and spacecraft and related machinery<br>32.5 Manufacture of medical and dental instruments and supplies | 50 Water transport,<br>51 Air transport<br>58 Publishing activities,<br>59 Motion picture, video and television programme production, sound recording and music publishing activities,<br>60 Programming and broadcasting activities,<br>61 Telecommunications,<br>62 Computer programming, consultancy and related activities,<br>63 Information service activities<br>64 to 66 Financial and insurance activities<br>69 Legal and accounting activities,<br>70 Activities of head offices; management consultancy activities,<br>71 Architectural and engineering activities; technical testing and analysis,<br>72 Scientific research and development,<br>73 Advertising and market research,<br>74 Other professional, scientific and technical activities,<br>75 Veterinary activities<br>78 Employment activities<br>80 Security and investigation activities<br>84 Public administration and defence, compulsory social security<br>85 Education<br>86 to 88 Human health and social work activities,<br>90 to 93 Arts, entertainment and recreation<br><br>Of these sectors, 59 to 63, and 72 are considered *high-tech services*. |

Sources: Eurostat/OECD (2011); cf. Laafia (2002, p. 7) and Leydesdorff *et al*. (2006, p. 186).

---

[2] Firms are classified in the ASIA database using ATECO 2007 codes, the Italian version of NACE Rev. 2.
[3] A complete index of NACE codes can be found, for example, at http://www.cso.ie/px/u/NACECoder/Index.asp .

We will additionally analyze the subsets of high- and medium-tech companies, and (high-tech) knowledge-intensive services, because one can expect very different dynamics for these sectors in contributing to synergy in the knowledge base of regions.

**5. Results**

*5.1. Regions*

Figure 3 provides a visualization of the percentage contribution of the twenty regions to the national synergy of Italy in 2015. The visualizations for 2008 and 2011 are not essentially different. The rank-order correlations among the regions in these three years are significantly the same (Spearman's $\rho > .99$; $p < 0.001$).

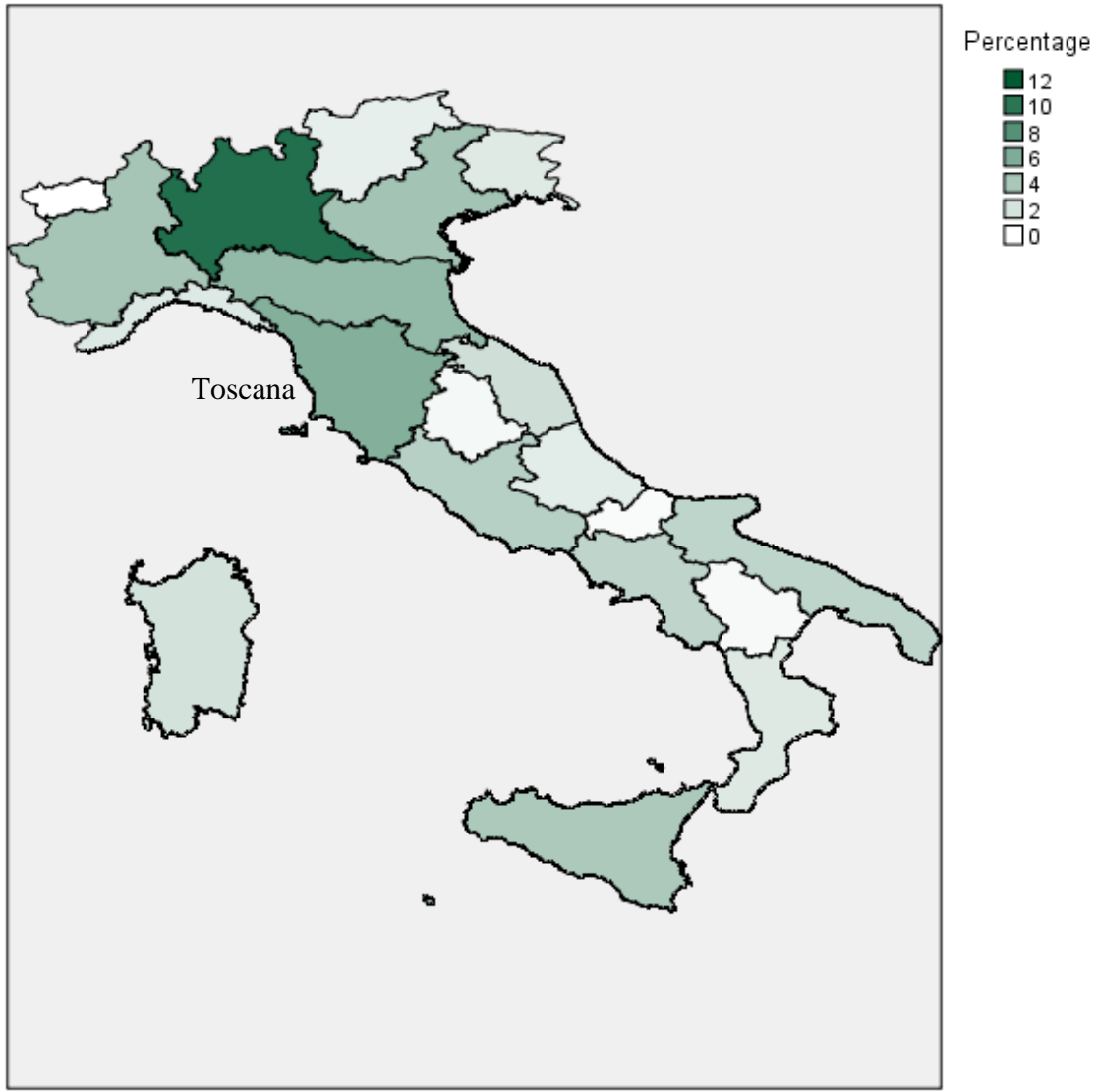

**Figure 3**: Percentages of contributions of the regions to the national synergy of Italy in 2015.

Figure 3 shows that Tuscany belongs to the northern part of Italy; the distinction of Central Italy including Tuscany is not supported by this data. Mountainous regions both along the Alps and in the Apennines are weakest in generating synergy. However, one should keep in mind that Italy has a system of excellent highways and trains that cross these regions. Their relative marginality is thus not likely to be due to the mountainous character of these regions, but a consequence of their structural positions.

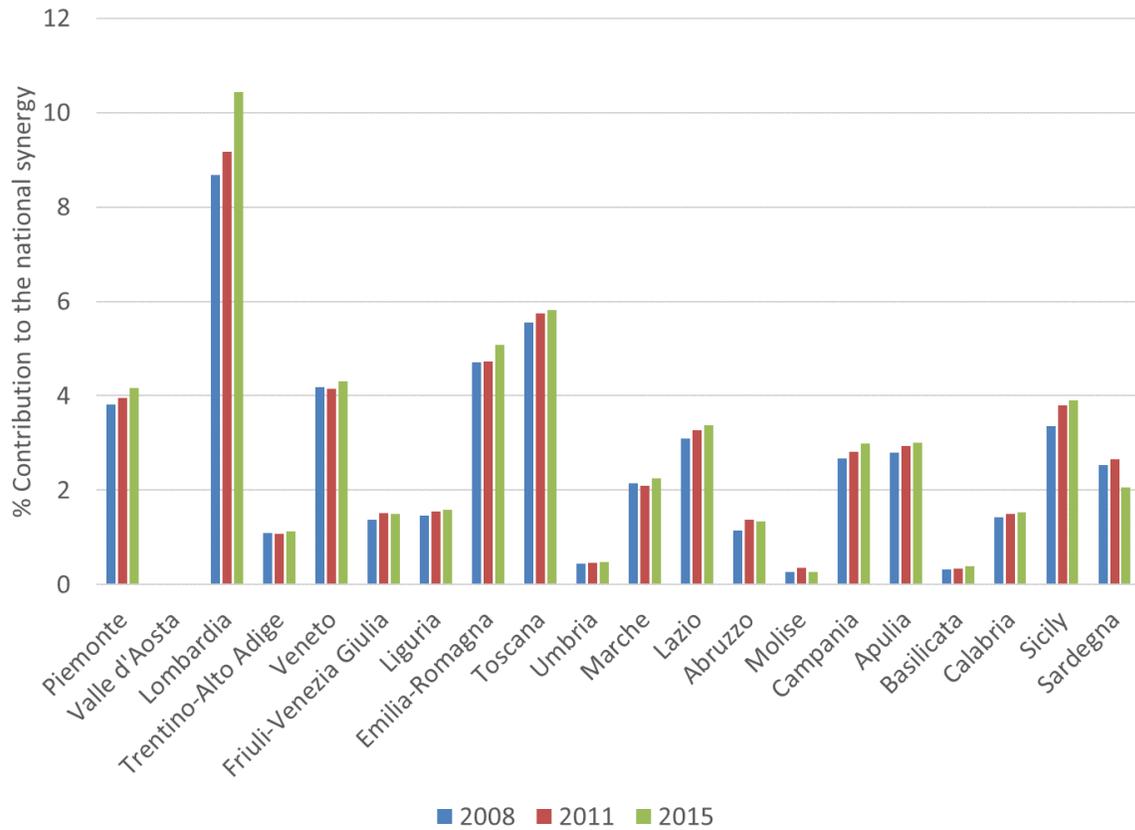

**Figure 4**: Percentages of contributions of the regions to the national synergy of Italy in 2008, 2011, and 2015.

Figure 4 shows that the triple-helix synergy increased over time in virtually all regions (but not in Sardegna). The strongest regions became even stronger in terms of their contributions to the national synergy. For example, Lombardia increased its leading contribution to the national synergy by another 1.8%. The percentage of synergy generated above the regional level—that is, the complement to 100% of the sum of the regional contributions—declined accordingly from 48.9% in 2008 to 44.4% in 2015 (– 4.5%). This reduction of above-regional synergy contribution over time as a percentage is consistent with the progressive withdrawal of innovation policy-making at the national level, and the growing importance of the devolved regions.

| Region | 2008 | 2011 | 2015 |
|---|---|---|---|
| Piemonte | 3.82 | 3.95 | 4.17 |
| Valle d'Aosta | 0.00 | 0.00 | 0.00 |
| Lombardia | 8.67 | 9.18 | 10.43 |
| Trentino-Alto Adige | 1.09 | 1.08 | 1.13 |
| Veneto | 4.19 | 4.15 | 4.31 |
| Friuli-Venezia Giulia | 1.37 | 1.51 | 1.49 |
| Liguria | 1.47 | 1.56 | 1.58 |
| Emilia-Romagna | 4.71 | 4.73 | 5.08 |
| Toscana | 5.55 | 5.75 | 5.81 |
| Umbria | 0.45 | 0.46 | 0.48 |
| Marche | 2.14 | 2.10 | 2.26 |
| Lazio | 3.09 | 3.27 | 3.38 |
| Abruzzo | 1.15 | 1.37 | 1.33 |
| Molise | 0.27 | 0.35 | 0.26 |
| Campania | 2.67 | 2.82 | 2.99 |
| Apulia | 2.79 | 2.94 | 3.01 |
| Basilicata | 0.32 | 0.33 | 0.38 |
| Calabria | 1.43 | 1.50 | 1.54 |
| Sicily | 3.36 | 3.79 | 3.89 |
| Sardegna | 2.54 | 2.66 | 2.07 |
| $T_0$ | 48.91 | 46.48 | 44.40 |

**Table 5**: Percentages of contributions of the regions to the national synergy of Italy in 2015.

In summary: regions become more important; but only 55% of the synergy is realized at the regional level. The other 45% is realized at the above-regional level (such as NUTS1, across the North/South divide, or in Italy as a whole).

*5.2. Northern, Central, and Southern Italy*

Using the classification of regions into Northern, Southern, and Central Italy as provided in Figure 1 above, Figure 5 shows the above-regional synergy development using three and two classes of regions, respectively, on the right side, and the values of $T_0$ on the basis of twenty regions on the left side. As noted, the latter declines from 48.9 to 44.4 %. The above-regional synergy development among the three groups of regions (north-south-center) is of the order of 22.5%, but is not consistently increasing as the supplement of the synergy among the 20 regions. Among two groups of regions (north-south) $T_0$ is further reduced to 18.2% in 2015.

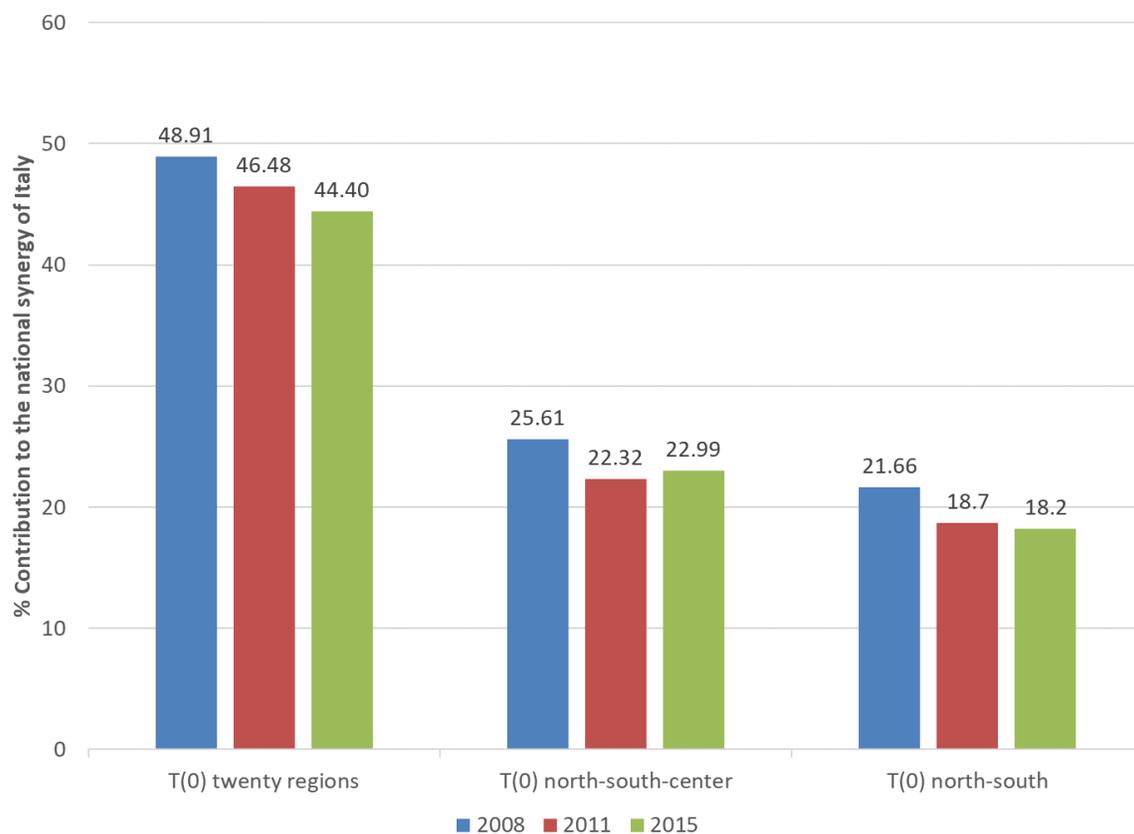

**Figure 5**: Above-regional synergy for Italy
on the basis of 20 NUTS2-regions (left) and three macro-regions (north – south – center).

In other words: if Tuscany is placed in the northern and not the southern part of the country, the northern part accounts for 47.0% of the synergy and the southern part for 34.9% with only 18.2% synergy at the national level. Both the northern and southern parts are more synergetic when compared with the division into three parts. Furthermore, values around 20% for the national surplus synergy were also found for other countries in previous studies. Adding Tuscany, which itself contributes only 5.8% to the synergy at the national level, to the northern part (instead of the central one), furthermore, increases the contribution of the north by more than 9% (= 46.95 – 37.90; in Table 6). Thus, an additional synergy is indicated by using this model of Italy.

Table 6: Percentage contributions of Northern, Southern, and Central Italy to the national synergy in 2015.

|  | *north-central-south* | *north-south* |
|---|---|---|
| *North* | 37.90 | 46.95 |
| *Center* | 17.50 | |
| *South* | 21.62 | 34.85 |
| Sum | 77.02 | 72.80 |
| $T_0$ | 22.98 | 18.20 |
| | 100 | 100 |

The conclusion is that considering Italy as twenty regions leaves 45% of the synergy in the Italian innovation system unexplained. This is extremely high when compared with other nations. In the USA, we found that the additional synergy at the national (above-state) level is only 2.8%. This is much less than we found in previous studies of national innovation systems: Norway (11.7%), China (18.0%), the Netherlands (27.1%), Sweden (20.4%), and Russia (37.9%). Italy scores above the Russian Federation when considered in these terms, but for very different reasons (Leydesdorff, Perevodchikov, & Uvarov, 2015). The high surplus in Russia is caused by the centralized nature of this system, while in Italy, the high surplus is unexplained because the wrong model is used for the country. When Italy is conceptualized as a country with two or three innovation systems, this description accords with those for other EU nations.

*5.3. Sectorial decomposition*

Using the NACE codes in Table 4, we can repeat the analysis for subsets of firms which are classified as high- or medium-high-tech, and knowledge-intensive services. Figures 6A and 6B show the distribution of the synergy for these subsets over the twenty regions. Of the approximately 4.3 million firms, 1,294,874 (29.8%) provide knowledge-intensive services, while only 40,083 (0.9%) are classified as MHTM in 2015. However, the differences between the distribution of the set and the subsets are marginal. Table 7 shows the rank-order correlations which are all above .95 ($p<.001$). In other words: both medium-high-tech and knowledge-intensive services are distributed proportionally over the country in terms of numbers of firms. Table 8 provides a summary of the results, including the values for these subsets as percentages of synergy in the two right-most columns.

**Table 7**: Rank-order correlations between the samples of firms classified as high- and medium-high-tech manufacturing (MHTM) and knowledge-intensive services (KIS) over the twenty regions of Italy.

|  |  |  | Full set | MHTM | KIS |
|---|---|---|---|---|---|
| Spearman's rho | Full set | Correlation Coefficient | 1.000 | .955** | .982** |
|  |  | Sig. (2-tailed) | . | .000 | .000 |
|  |  | N | 20 | 20 | 20 |
|  | MHTM | Correlation Coefficient | .955** | 1.000 | .950** |
|  |  | Sig. (2-tailed) | .000 | . | .000 |
|  |  | N | 20 | 20 | 20 |
|  | KIS | Correlation Coefficient | .982** | .950** | 1.000 |
|  |  | Sig. (2-tailed) | .000 | .000 | . |
|  |  | N | 20 | 20 | 20 |

**. Correlation is significant at the 0.01 level (2-tailed).

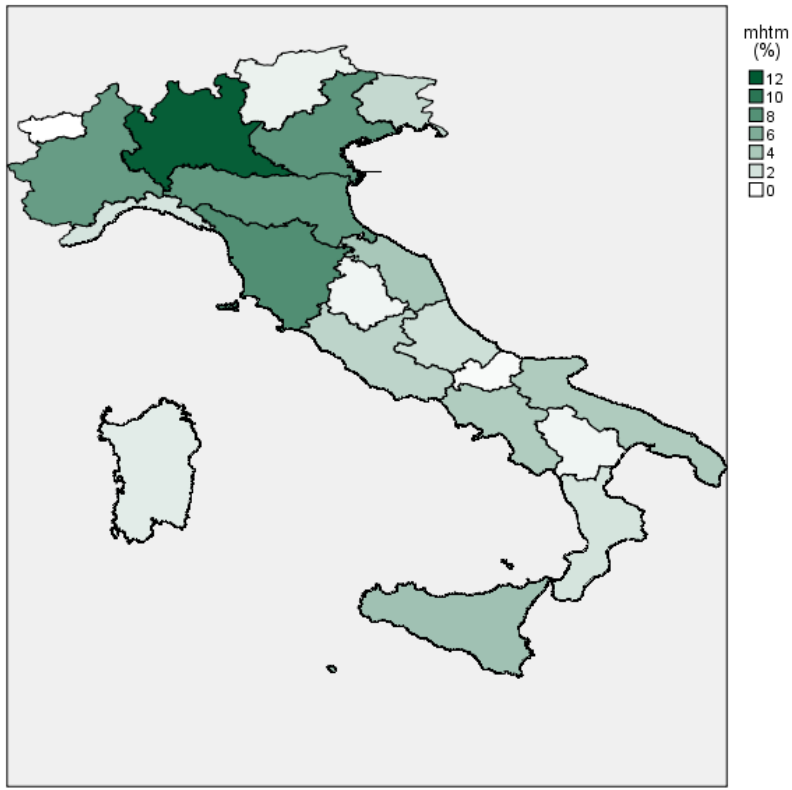 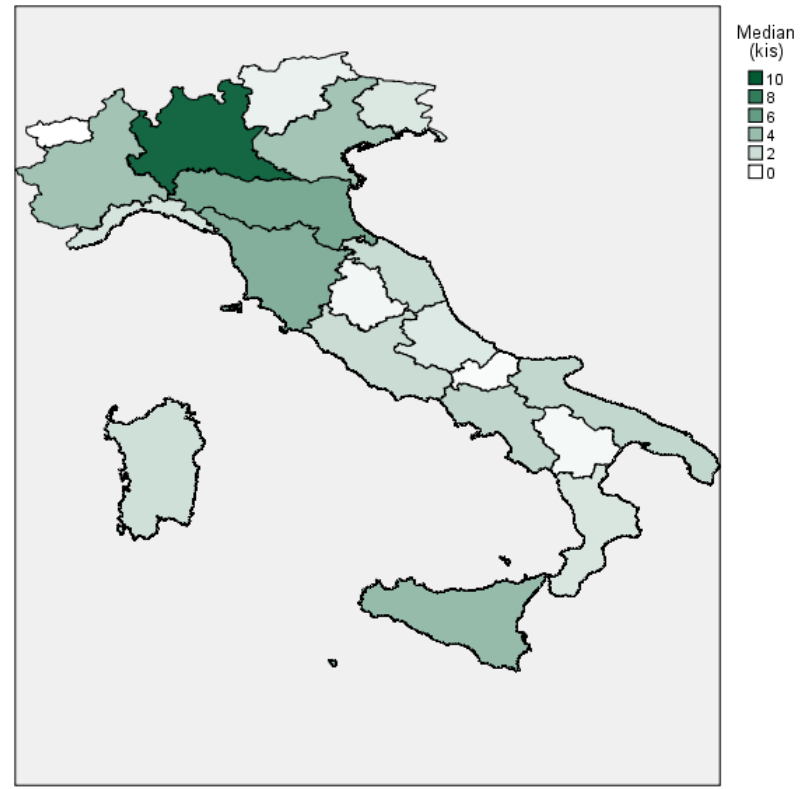

**Figure 6A**: Regional decomposition of the synergy in the Italian innovation system for medium- and high-tech companies

**Figure 6B**: Regional decomposition of the synergy in the Italian innovation system for knowledge-intensive services.

**Table 8**: Summary table of percentages of contributions to the synergy
in the Italian innovation system (2015)

| Region | 2015 | north_south_center | north_south | MHTM | KIS |
|---|---|---|---|---|---|
| *Piemonte* | 4.17 | 37.90 | 46.95 | **7.14** | 3.58 |
| *Valle d'Aosta* | 0.00 | | | 0.00 | 0.00 |
| *Lombardia* | 10.43 | | | 11.68 | **9.19** |
| *Trentino-Alto Adige* | 1.13 | | | 0.94 | 0.80 |
| *Veneto* | 4.31 | | | **7.66** | 3.54 |
| *Friuli-Venezia Giulia* | 1.49 | | | 2.72 | 1.38 |
| *Liguria* | 1.58 | | | 1.93 | 1.59 |
| *Emilia-Romagna* | 5.08 | | | **7.40** | 5.20 |
| *Toscana* | 5.81 | 17.50 | | **8.15** | 4.81 |
| *Umbria* | 0.48 | | 34.85 | 0.67 | 0.50 |
| *Marche* | 2.26 | | | 4.06 | 2.12 |
| *Lazio* | 3.38 | | | 3.07 | **2.07** |
| *Abruzzo* | 1.33 | 21.62 | | 2.32 | 1.30 |
| *Molise* | 0.26 | | | 0.30 | 0.21 |
| *Campania* | 2.99 | | | 3.70 | 2.45 |
| *Apulia* | 3.01 | | | 3.76 | 2.36 |
| *Basilicata* | 0.38 | | | 0.70 | 0.42 |
| *Calabria* | 1.54 | | | 1.96 | 1.47 |
| *Sicily* | 3.89 | | | 4.44 | 4.09 |
| *Sardegna* | 2.07 | | | 1.34 | 1.85 |
| Sum | 55.60 | 77.01 | 71.80 | 73.94 | 48.93 |
| $T_0$ | 44.40 | 22.99 | 18.20 | 26.06 | 51.07 |

We boldfaced in Table 8 some values in the right-most columns for regions with outlier values for MHTM and/or KIS. Piemonte, Veneto, Emilia-Romagna, and Toscana have contributions to the synergy when we focus on MHTM more than two percent higher than without this focus. Lombardia, Marche, and Friuli-Venezia Giulia follow with more than one percent higher values.

Unlike manufacturing, services can be offered nation-wide or even beyond the nation, and thus tend to uncouple from the location, leading to a negative effect on the local synergy. In Italy, this is the case mainly for services in Lombardia and Lazio, while these two regions contain the two metropoles of Milano and Rome with airports, etc. Toscana (Florence) and Veneto (Venice) follow with smaller effects.

In Southern Italy, there are no effects from either MHTM or KIS. A negative effect of MHTM is indicated for Lazio, probably meaning that some manufacturing may have the administrative offices in Rome without contributing to the knowledge-based synergy in this region. Sardegna also has such a negative effect when focusing on MHTM because these sectors are marginal in the local economy.

**6. Conclusions and discussion**

*6.1. Summary*

Innovation systems are not *a priori* bound by administrative and political borders. In analogy to "national innovation systems" (Freeman, 1987; Lundvall, 1988, 1992; Nelson, 1993), many studies have argued for studying "regional innovation systems" such as Wales or Catalonia (Braczyk, Cooke, & Heidenreich, 1998; Cooke, 1998, 2002). In our opinion, one should not make the choice between studying regions or nations on normative grounds and across the board. The function of regions in an otherwise relatively homogeneous country (e.g., France or Denmark) is different from that in a country with a federal structure, such as Belgium.

From this perspective, Italy is an interesting case because there is a traditional divide between the north and the south, but there are also common denominators such as a single language (with small exceptions), national institutions such as state universities a national research council (CNR) with a similar structure in all regions, and a central government without a federal structure. Regions have become more important during the last two decades, because of the devolution policies of the central government and the emphasis on regions in EU policies.

One would expect the coherence of an innovation system to be a mixture of both national and regional aspects. The research question then becomes: how much innovation-systemness is generated at the various levels? How is this innovation-systemness distributed and specialized in specific regions? The synergy measure developed in this paper enables us to address these questions empirically.

Italy as a nation is integrated, albeit not only at the level of the twenty regions. Eight regions in Northern Italy (including Tuscany) are well developed as innovation systems. Taken together (Table 5), these eight regions contribute only 34.0% to the national synergy. However, as a separate subsystem Northern Italy contributes 47.0% of the synergy (Table 6). This is 13% more than the sum of the individual regions. The regions on the Northern borders with different cultural orientations (Alto-Adige and Valle d'Aosta) contribute marginally to the synergy in the Italian system.

If we apply the same reasoning to Southern Italy (the Italian *Mezzogiorno*), twelve regions contribute 21.6% to the national synergy. Considered as a subsystem (Table 6), the South contributes 34.9%; that is, another 13.3% more synergy. On top of these two sub-systems, Italy as a nation contributes 18.2% to the national synergy. Thus, most synergy is found by considering Italy in terms of a northern and southern part, with Tuscany as part of Northern Italy.

The division in terms of North, South, and Central Italy relocates Tuscany into the central part. Using this model, Central Italy improves its 11.9% contribution to the national synergy to 17.5%—that is, +5.6%—while Southern Italy in this configuration improves its contribution from 15.5 to 21.6%, that is 6.1%. The seven remaining regions of Northern Italy in this case generate 28.2% of the synergy as an aggregate, but 37.9% as a single system. The additional synergy is now 9.7% and thus much less than the 13% generated additionally in the configuration of only North and South. Both Northern and Southern Italy (including Central Italy) perform better as innovation systems in terms of synergy generation in a configuration of two sub-systems. The difference is of the order of 5% synergy.

As one would expect, synergy is enhanced by focusing on high- and medium-tech manufacturing. Rome and Milano function as metropolitan centers of innovation systems, followed by Florence and the Venice region (including the harbour). Unlike Spain, where Barcelona and Madrid function as metropolitan innovation systems without much further integration into the remainder of the country (Leydesdorff & Porto-Gómez, 2018), the Italian system is integrated also in terms of MHTM and KIS.

*6.2. Policy implications*

If innovation policy is focused on the regional level, one may miss important opportunities in inter-regional interactions. In other words, the coordination of innovation policies among regions, particularly within each of the two major innovation (sub)systems of Italy, would be desirable. More generally, our results provide further support for the argument that administrative borders which originated for historical and administrative reasons should be examined critically in terms of their functionality for innovation systems, particularly in a knowledge-based economy which is far more networked than a political economy (Leydesdorff, Ivanova, & Meyer, 2018; forthcoming).

The knowledge dynamics added to the economic and political dynamics generates a complex system with a volatile dynamics that tends to self-organize its boundaries (Bathelt, 2003). A complex system is resilient and thus adapts to signals that do not accord with its internal dynamics. A political administration that is not reflexively aware of and informed about how the relevant innovation systems are shaped, may miss the requisite variety to steer these systems and feel overburdened by the unintended consequences of its actions (Ashby, 1958; Luhmann, 1997).

*6.3. Limitations and future perspectives*

One limitation of this study remains the nature of the data. The current statistics tend to attribute a single address (for example, headquarters) to firms with multiple locations. In this study, we also used only the first NACE code of each firm. The possibility to search for optima in a phase space of the three (or more) distributions may reveal growth potentials of combinations that have remained hitherto unnoticed.

**Acknowledgment**

We thank Federica Rossi for her comments and contributions.